\documentclass[aps,prd,twocolumn,superscriptaddress,showpacs,eqsecnum,floats]{revtex4}

\usepackage{graphicx}

\begin{document}

\preprint{OU-TAP 271}

\title{Results of the search for inspiraling compact star binaries \\
from TAMA300's observation in 2000-2004}

\author{Tomomi Akutsu}
\affiliation{Institute for Cosmic Ray Research, The University of Tokyo, Kashiwa, Chiba 277-8582, Japan} 
\author{Tomotada Akutsu}
\affiliation{Department of Astronomy, The University of Tokyo, Bunkyo-ku, Tokyo 113-0033, Japan} 
\author{Masaki Ando}	
\affiliation{Department of Physics, The University of Tokyo, Bunkyo-ku, Tokyo 113-0033, Japan} 
\author{Koji Arai}	
\affiliation{National Astronomical Observatory, Mitaka, Tokyo 181-8588, Japan} 
\author{Akito Araya}	
\affiliation{Earthquake Research Institute, The University of Tokyo, Bunkyo-ku, Tokyo 113-0032, Japan} 
\author{Hideki Asada}	
\affiliation{Faculty of Science and Technology, Hirosaki University, Hirosaki, Aomori 036-8561, Japan} 
\author{Youichi Aso}	
\affiliation{Department of Physics, The University of Tokyo, Bunkyo-ku, Tokyo 113-0033, Japan} 
\author{Mark A. Barton}	
\affiliation{Institute for Cosmic Ray Research, The University of Tokyo, Kashiwa, Chiba 277-8582, Japan} 
\author{Peter Beyersdorf}	
\affiliation{National Astronomical Observatory, Mitaka, Tokyo 181-8588, Japan} 
\author{Youhei Fujiki}	
\affiliation{Faculty of Science, Niigata University, Niigata, Niigata 950-2102, Japan} 
% ...

\collaboration{The TAMA Collaboration}
\noaffiliation


\begin{abstract}

We analyze the data of TAMA300 detector to search for 
gravitational waves from 
inspiraling compact star binaries with masses of the component stars 
in the range $1-3M_\odot$. 
In this analysis, 2705 hours of data, taken during the
years 2000-2004, are used for the event search. 
We combine the results of different observation runs, and obtained 
a single upper limit on the rate of 
the coalescence of compact binaries in our Galaxy of 20 per year at a 90\% confidence level. 
In this upper limit, the effect of various systematic errors such like the uncertainty of 
the background estimation and the calibration of the detector's sensitivity are included. 
\end{abstract}


\maketitle

\section{Introduction}

Several laser interferometric gravitational wave detectors of the first generation
have been operated to detect gravitational wave. 
These include GEO\cite{ref:GEO}, LIGO\cite{ref:LIGO}, TAMA300\cite{ref:TAMA}, 
and VIRGO\cite{ref:VIRGO}. 
These detectors have been improved their sensitivity very rapidly in the past 
several years. 
The direct detection of gravitational waves is important not only because it will become
a new astronomical tool to observe our universe, but also because
it will become a new tool to test general relativity and other gravity theories
in a strong gravity field region.  

In this paper, we present results of the data analysis of the TAMA300 detector
to search for gravitational waves produced by inspiraling compact star binaries, 
comprised of non-spinning neutron stars and/or black holes. 
Inspiraling compact binaries 
are considered to be one of the most promising sources for ground based 
laser interferometers. 
TAMA300 has been performed nine observation runs of the detector since 1999. 
The total amount of data is more than 3000 hours. 
Given such large amount of data, it is very interesting to analyze the data to search 
for candidate gravitational wave events and to set an upper limit to the event rate. 

In the past, there were several works which searched for inspiraling compact binaries
using laser interferometer data. 
Allen {\it et al.} \cite{ref:40m} analyzed LIGO-40m's data in the mass range 
$1-3M_\odot$, and obtained an upper limit of 0.5 [1/hour] on the Galactic 
event rate.  
The data from "Data Taking 2" (DT2) of TAMA300  in 1999 was analyzed 
in the mass range $0.3-10M_\odot$ \cite{ref:TAMA_DT2}, 
and an upper limit of 0.59 [1/hour] on the event rate with signal-to-noise ratio
greater than 7.2 was obtained. 
TAMA300's DT6 data and LISM-20m's data taken in 2001 were analyzed to search for 
coincident signals, and obtained an upper limit of 
0.046 [1/hour] on the nearby event rate within 1kpc from the Earth \cite{ref:TAMA_LISM}. 
Abbott {\it et al.} \cite{ref:LIGO_S1} analyzed LIGO's "1st Scientific run" (S1) data taken in 2002, 
and obtained an upper limit of 1.7$\times 10^2$ per year per Milky Way 
Equivalent Galaxy (MWEG) in the mass range $1-3M_\odot$. 
Abbott {\it et al.} analyzed LIGO S2 data taken in 2003, and obtained an upper limit
of 47 per year per MWEG in the mass range $1-3M_\odot$ \cite{ref:LIGO_S2N}, 
63 per year per Milky Way halo in the mass range $0.2-1M_\odot$ \cite{ref:LIGO_S2Ma}. 
LIGO's S2 data was also analyzed to search for binary black hole inspirals 
in the mass range $3-20M_\odot$, and an upper limit of
37 per year per MWEG was obtained \cite{ref:LIGO_S2B}. 
LIGO's S2 data and TAMA300's DT8 data were analyzed to search for coincident 
signals and an upper limit of 49 per year per MWEG was obtained \cite{ref:LIGO_TAMA}. 
In all of above cases, there were no signals that could be identified as gravitational 
waves. 

In this paper, we analyze the data from DT4, DT5, DT6, DT8, and DT9 
of TAMA300. A part of DT6 data which was coincident with LISM 
was already analyzed in \cite{ref:TAMA_LISM}.  
The initial results of the analysis of DT8 data was reported in Ref. \cite{ref:DT8}. 
A part of DT8 data which was coincident with LIGO S2 
was already analyzed in \cite{ref:LIGO_TAMA}. 
In this paper, we analyze these data again together with the other data in a unified way. 
Until DT6 observation, TAMA300 was the only large scale laser interferometer 
which was operated. Thus, it is important to analyze such data to search for possible gravitational wave 
signals.  Further, in order to take advantage of the long length of data from DT6, DT8 and DT9, 
we combine the results from DT6, DT8 and DT9 data
and obtain a single upper limit on the rate of 
the coalescence of compact binaries in our Galaxy.
We also evaluate the systematic errors caused by the uncertainty of the calibration 
and the background trigger rate. Other errors such like the uncertainty of the distribution model of sources, 
and the uncertainty of the theoretical templates are also evaluated. 
These systematic errors are taken into account to evaluate the upper limit. 

This paper is organized as follows. In section II, we overview the detector and the 
data we analyze. In section III, the analysis method is presented. 
In section IV, the results of the analysis are presented. In section V, 
the evaluation of the detection probability of the Galactic signals and 
the upper limit to the event rate are shown. 
In section VI, we evaluate the errors due to various error sources, and its effect
to the upper limit. 
In section VII, we summarize the results and present the conclusion.

\begin{table*}[t]
\renewcommand{\arraystretch}{1.5}
\begin{tabular}{c c c c c}
\hline \hline
   & Period & Typical strain noise & Observed data & Analyzed data \\
   &        & [$1/\sqrt{\rm{Hz}}$] & [hours] & [hours] \\
\hline 
DT1 & 6-7 Aug. 1999 		& $3 \times 10^{-19}$   & 11 & - \\
DT2 & 17-20 Sept. 1999 		& $3 \times 10^{-20}$   & 31 & - \\
DT3 & 20-23 April 2000		& $1 \times 10^{-20}$   & 13 & - \\
DT4 & 21 Aug.-4 Sept. 2000	& $1 \times 10^{-20}$   & 154.9 & 147.1\\
DT5 & 2-10 Mar. 2001		& $1.7 \times 10^{-20}$ & 107.8 & 95.26\\
DT6 & 1 Aug.-20 Sept. 2001	& $5 \times 10^{-21}$   & 1049 & 876.6\\
DT7 & 31 Aug.-2 Sept. 2002	& 					& 25 & -\\
DT8 & 14 Feb.-14 April 2003	& $3 \times 10^{-21}$ &  1163 & 1100\\
DT9 & 28 Nov. 2003 - 10 Jan. 2004& $1.5 \times 10^{-21}$ & 556.9 & 486.1\\
\hline
Total & & & 3111.6 & 2705 \\
(DT6,8,9 for upper limit) & & & & (2462.8) \\
\hline \hline
\end{tabular} 
\caption{\label{tab:runlist} 
Observation history of TAMA300. For each observation run, the period of the observation, 
typical strain equivalent noise level 
around the most sensitive frequency region, the length of data observed, and the the length
of data analyzed in this paper are shown. }  
\end{table*}

\section{Data from the TAMA300 detector}

TAMA300 is a Fabry-Perot-Michelson interferometer with baseline length of 300m
located in Mitaka, Tokyo ($35^{\circ}40^{\prime}$N, $139^{\circ}32^{\prime}$E).  
The history of the observation run of TAMA300 is listed in Table \ref{tab:runlist}. 
Until DT6, the detector was operated without the power recycling system. 
After DT6, the power recycling system was installed. 
The main signal of detector is recorded with a 20 kHz, 16 bit data-acquisition system. 
There are more than 150 signals which monitor the condition of the detector, 
and the environment of detectors. 
During the operation, the mirrors of the detector are shaken by a 625 Hz sinusoidal signal
in order to calibrate the detector sensitivity continuously. 

We use DT4, DT5, DT6, DT8, and DT9 data of TAMA300. 
The observations of TAMA300  were interrupted by the unlock of the detector.
They were sometimes suspended manually for maintenance.  
By removing such dead time, the total length of data available for 
the data analysis is 3032 hours.  
Among them, we do not use the first 6.5 minutes of data 
just after the detector recovers from the dead time, 
because such data often contain signals due to the excitation of 
the violin modes of pendulum wires, 
and/or other signals caused by disturbance during the dead time. 
The data from the detector are transfered into the strain equivalent data 
by applying the transfer function. 
The fluctuation of the transfer function at each time is determined 
by computing the optical gain. 
We do not use the data if the value of the optical gain deviates from the 
average value significantly. 
The total amount of data remained after removing such bad quality parts 
are 2705 hours. 
We analyze these data to search for gravitational wave events.  
However, we do not use DT4 and DT5 data to set the upper limit 
for the event rate, because the length of data 
from these runs are much shorter than DT6-8-9, and because 
quality of data of these runs are much worse than those of DT6-8-9. 
The total amount of data used for setting the upper limit is 2462.8 hours.

\section{Analysis method}

The standard method to search for gravitational wave signals with known wave forms
in noisy data is the matched filtering method, in which 
we search for the best matched parameters of the theoretical wave form
by cross-correlating the data with the theoretical wave form. 
As the theoretical wave form, 
we use the non-spinning, restricted post-Newtonian (PN) wave form in which the phase is given to 
high post-Newtonian order, but only the leading quadrupole term is contained 
in the amplitude. We use the phase formula derived from 2.5 PN approximation. 
Although the current best formula by the PN approximation is the 3.5 PN formula 
\cite{ref:PN}, the error due to the use of 2.5 PN formula instead of 3.5PN formula is small
for binaries of mass considered in this paper (see Section IV). 
On the other hand, the PN approximation itself may
contain errors due mainly to the relativistic effects in the region when the 
orbital radius is the same order as the gravitational radius of stars. 
These effects will be incorporated in the systematic error
to the detection probability of signals. 

The basic formula of the matched filtering method  is given by
\begin{eqnarray}
\rho&=&\sqrt{(s,h_0)^2+(s,h_{\pi/2})^2}, \label{eq:defrho}
\end{eqnarray}
where
\begin{eqnarray}
(a,b)&\equiv& 2\int^{\infty}_{-\infty} \frac{\tilde{a}^*(f) \tilde{b}(f)}{S_n(|f|)} df,
\label{eq:definner}
\end{eqnarray}
and where $\tilde{a}(f)$ and $\tilde{b}(f)$ are the Fourier transformation of time sequential data, 
$a(t)$ and $b(t)$. 
The Fourier transformation is defined by 
\begin{eqnarray} 
\tilde{a}(f)=\int^{\infty}_{-\infty} a(t) e^{2\pi i f t} dt.
\end{eqnarray}
The function $s(t)$ is the time sequential data from the detector. 
Two functions, $h_0$ and $h_{\pi/2}$, are the templates in the 
frequency domain, which are normalized as $(h_0,h_0)=(h_{\pi/2},h_{\pi/2})=1$.
The Fourier transformation of them, $\tilde{h}_0(f)$ and $\tilde{h}_{\pi/2}(f)$,
are computed by the stationary phase approximation. 
We thus have the orthogonality of them, i.e., $(h_{0}, h_{\pi/2})=0$. 
The function $S_n(f)$ is the one-sided power spectrum density of noise of the detector. 

The parameters which describe the wave form are the time of coalescence, $t_c$, 
the phase of wave at the coalescence, $\phi_c$, the total mass $M\equiv m_1+m_2$ 
and the non-dimensional reduce mass $\eta\equiv m_1 m_2/M^2$ of the binary. 
We search for the parameters which give the maximum of $\rho$. 
In the formula (\ref{eq:defrho}), the maximization over the phase 
is already taken analytically. The value of parameters, $t_c$, $M$ and $\eta$, 
which maximize $\rho$ are searched numerically. 


The data are divided into subsets of data with length 52.4 seconds. 
Each subset of data has overlapping data with adjacent data for 4.0 seconds. 
Each subset of data is Fourier transformed, and  
the components more than 5kHz are removed. 
The data are converted to the strain 
equivalent data $\tilde{s}(f)$ by the transfer function.  
The power spectrum density of noise $S_n(f)$ is evaluated at neighbor of
each subset. Details of the method to evaluate $S_n(f)$ was described in 
\S III.B. of Ref. \cite{ref:TAMA_LISM}.
With the subset of data, we compute $\rho$. 
For each small time interval with length $\Delta t_c=25.6$msec, we search for 
$t_c, M$ and $\eta$ which give the maximum of $\rho$. 
The value of $\rho$ at all of $t_c$ 
can be computed automatically from the inverse FFT of 
the inner product, Eq.(\ref{eq:definner}), with respect to $t_c$.
The search for the best matched $M$ and $\eta$ is done by introducing the grid points 
in the two dimensional mass parameter space. 
The range of masses of each member star of binaries is 
set to 1 to 3 $M_\odot$. The grid separation length 
is determined so that the minimal match is less than 3\% \cite{ref:Owen}. 
The actual mass parameters we use for setting up the mass parameter space
are those discussed in \cite{ref:TT}. 
We define {\it a trigger} by the local maximum of $\rho$ 
in each small time interval with length, $\Delta t_c=25.6$msec, 
and in the whole mass parameter region, 
together with the parameters, $t_c$, $M$ and $\eta$ which realize the local maximum. 

\section{Trigger distributions}


The data of TAMA300 contain non-stationary, non-Gaussian noise. 
Such noise cause many triggers with rate much larger than 
that expected in the stationary Gaussian noise. 
In order to distinguish such spurious triggers from triggers caused by real 
gravitational wave signals, 
we compute $\chi^2$ value for each trigger with $\rho\geq 7$. 
The definition of $\chi^2$ can be found in \cite{ref:Allen}
\footnote{This $\chi^2$ means the reduced chi square, which 
is the usual chi square divided by the degree of freedom.
In this paper, the number of bins of frequency region is taken to be 16.  
Thus, the degree of freedom of $\chi^2$ is 30. }. 
This $\chi^2$ is defined such that it is independent from the amplitude
of signal if the wave form of the signal and the template are identical.
However, since our template parameters are defined discretely, 
and thus the signals are different from the templates in general, 
when the amplitude of signal becomes larger, $\chi^2$ becomes larger. 
In order not to lose real signals with large $\chi^2$, we define $\zeta=\rho/\sqrt{\chi^2}$
as a new statistic \cite{ref:TAMA_LISM}. 
The statistic, $\zeta$, was used in our previous analysis \cite{ref:TAMA_LISM}, and was found to be
useful to distinguish the spurious triggers from triggers caused by real gravitational wave signals. 

\begin{figure}[htbp]
\includegraphics[width=7cm]{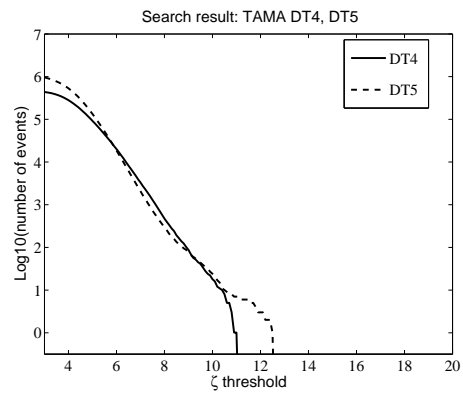}
\caption{\label{fig:dt45}The cumulative number of triggers as a function of $\zeta$ for DT4 and DT5.}
\end{figure}

\begin{figure}[htbp]
\includegraphics[width=7cm]{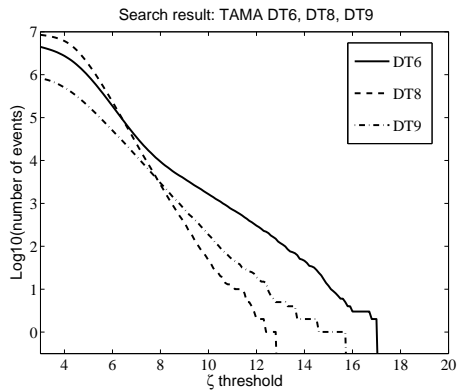}
\caption{\label{fig:dt689}The cumulative number of triggers as a function of $\zeta$ for DT6, DT8 and DT9.}
\end{figure}

The cumulative number distribution of triggers as a function of $\zeta$ for each observation run
are plotted in Fig.\ref{fig:dt45}-\ref{fig:dt689}.

In these plots, there are no triggers which deviate from the tail of the distribution 
of triggers significantly. 
This fact suggests that there is no candidate trigger which 
can be interpreted as real gravitational signal. 

\section{Upper limit to the Galactic event rate}

\begin{figure}[t]
\includegraphics[width=7cm]{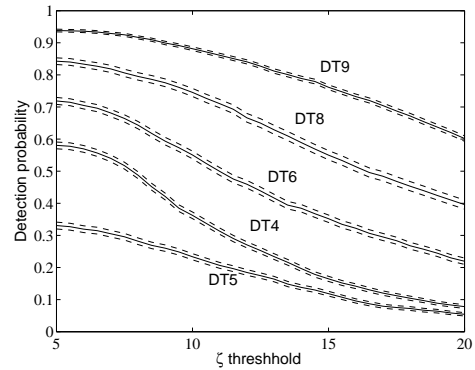}%
\caption{\label{fig:efficiency_all}Detection probability of Galactic binaries inspirals for each observation run.
The dashed lines show the uncertainty of the Monte Carlo simulations. }
\end{figure}

In this section, we evaluate the upper limit on the rate of the inspirals of compact binaries in our Galaxy. 
In order to do this, we first evaluate the detection probability of Galactic signals by adding the 
signals to the real data, and by analyzing the data by the same analysis pipeline used 
in the real analysis. 
We assume the distribution of compact star binaries in our Galaxy given by
\begin{eqnarray}
dN=e^{-r^2/(2r_0^2)}e^{-Z/h_z} r dr dZ,
\end{eqnarray}
where $r$ is the radius from the center of the Galaxy, $r_0=8.5$kpc, $Z$ is height from 
the galactic plane, and $h_z=1$kpc. 
We assume that mass of each component star is uniformly distributed between 1 to 3 $M_\odot$,
because we do not know much about the mass distribution model of binary compact stars
including black holes and/or neutron stars. 
We also assume uniformly distributed 
inclination angle of the orbital plane and the polarization angle of signals. 
The obtained detection probability is plotted in Fig. \ref{fig:efficiency_all}. 
The DT9's data are the most sensitive to the Galactic events. 
Actually, the detection probability by the second half of the DT9 data is much better than 
that by the first half data. The first half data of DT9 was not very stable. Many triggers with 
large $\zeta$ were produced by instrumental noise during that period. They degrade the average detection
probability in DT9. 

The upper limit to the event rate from each observation run, $R_i$ ($i=$ DT6, DT8, DT9), 
is derived by 
\begin{eqnarray}
R_i=\frac{N_i}{T_i\epsilon_i}, \quad 
\end{eqnarray}
where $T_i$ is the length of data, $\epsilon_i$ is the detection probability, and 
$N_i$ is the upper limit to the number of event derived by the following formula: 
\begin{eqnarray}
\frac{e^{-(N_i+N^{(i)}_{\rm bg})} \sum_{n=0}^{n=N^{(i)}_{\rm obs}}
\frac{(N_i+N^{(i)}_{\rm bg})^n}{n!}}
{e^{-N^{(i)}_{\rm bg}} \sum_{n=0}^{n=N^{(i)}_{\rm obs}}\frac{(N^{(i)}_{\rm bg})^n}{n!}}
=1-{\rm C.L.} \, ,
\label{eq:upperlimit}
\end{eqnarray} 
where $N^{(i)}_{\rm obs}$ is the observed number of triggers which exceed a threshold, 
$N^{(i)}_{\rm bg}$ is the number of triggers which are caused by noise alone, 
and C.L. is a confidence level. 

\begin{figure}[htbp]
\includegraphics[width=7cm]{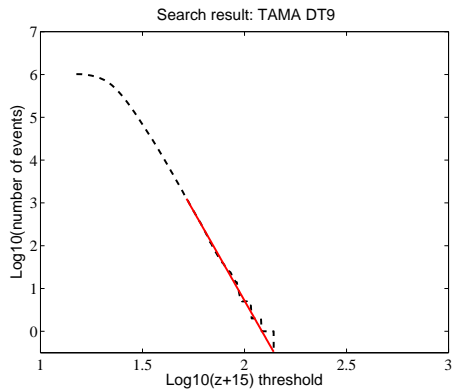}
\caption{\label{fig:fittingDT9} The dashed line denotes the cumulative number of 
triggers as a function of $\log_{10}(z+p-1)$,  $(p=16)$, for DT9. 
The solid line is the result of the least square fitting to this trigger distribution.}
\end{figure}

We set the false alarm rate to 1 event par year. 
The threshold which corresponds to this false alarm rate 
is evaluated by fitting the trigger distribution 
assuming that all triggers are caused by noise. 
We note that $z\equiv\zeta^2/2$ obeys the F distribution
with degree of freedom, $(2,2p-2)$, when the data are the Gaussian noise. 
Here, $p$ is the number of bins in the frequency region which is used to define $\chi^2$, 
and we set $p=16$.
In this case, the variable $z$ obeys the probability 
density function given by $(p-1)^p(z+p-1)^{-p}$. 
The cumulative number of triggers as a function of the threshold, $N(z)$, is 
proportional to $N(z)\propto (p-1)^{p-1}(z+p-1)^{-p+1}$. 
Thus, the plot of $\log N(z)$ - $\zeta$ is not linear, but 
$\log N(z)$ $-$ $\log(z+p-1)$ becomes linear.  
Although, TAMA300's data show non-Gaussian property, 
these facts suggest that $\log N(z)$ $-$ $\log(z+p-1)$ plot 
may be more suitable for an accurate evaluation of the false alarm rate
as a function of the threshold. 
We find that this is actually the case for DT8 and DT9. 
In Fig.\ref{fig:fittingDT9}, we show the result of the fitting for DT9 case. 
The thresholds obtained in this way are listed in Table \ref{tab:upperlimit}.
On the other hand, the same plot does not become linear in DT6 case. 
We then conservatively select the region of the fitting so that we have a larger
threshold for a given false alarm rate. 

With these thresholds, we have $N^{(i)}_{\rm obs}=0$ for all cases. 
From Eq.(\ref{eq:upperlimit}), the upper limit to the number of event is $N_i=2.3$
for a confidence level C.L.=90\%. 
We obtain the upper limit for the Galactic event rate, 
130 [yr$^{-1}$] from DT6, 30 [yr$^{-1}$] from DT8, and 60 [yr$^{-1}$] from DT9. 
The thresholds, the detection probability, and 
the upper limit for the Galactic event rate for each run 
are listed in Table \ref{tab:upperlimit}. 
The most stringent upper limit is obtained from DT8 data. 
This is because the length of data is the longest among three runs, and 
because the detection probability is comparable to that of DT9 on average. 

We combine these results and obtain a single upper limit. The upper limit 
from different observation runs is given by 
\begin{eqnarray}
R=\frac{N_{\rm UL}}{\sum_i T_i \epsilon_i},
\label{eq:combinedupperlimit}
\end{eqnarray}
where $N_{\rm UL}$ is the upper limit to the number of events derived by all of data.
We adopt the same threshold $\zeta^*$ for each run listed in Table \ref{tab:upperlimit}.
The total number of background triggers is $\sum_i N_{\rm bg}^{(i)}=0.281$. 
Thus, we have 1 event per year as a false alarm rate for combined DT6-DT8-DT9 data.  
Since the total number of triggers observed is zero, $\sum_i N^{(i)}_{\rm obs}=0$, 
we have $N_{\rm UL}=2.3$ for C.L.=90\%. 
From Eq. (\ref{eq:combinedupperlimit}), the combined upper limit to the event rate becomes 
\begin{eqnarray}
R=17  ~[{\rm yr}^{-1}].
\end{eqnarray}

\begin{table}[htbp]
\begin{tabular}{cccc}
\hline
& DT6 & DT8 & DT9 \\ \hline
{\small Observation time [hours]} & 876.6 & 1100 & 486.1 \\ \hline
{\small Threshold $\zeta^*$} & 21.8 & 13.7 & 17.7 \\ \hline
{\small  $N_{bg}^{(i)}$} & 0.1000 & 0.1255 & 0.0555 \\ \hline
{\small Detection probability} & $0.18$ & $0.60$ & $0.69$ \\ \hline
{\small $(\delta R_i)_{\rm fluct}$ [yr$^{-1}$]} & $+20.6$ & $+2.52$ & $+4.04$ \\ 
& $-24.0$ & $-2.82$ & $-3.77$ \\ \hline
{\small $(\delta R_i)_{\rm model}$ [yr$^{-1}$]} & $+55.4$ & $+4.18$ & $+6.84$ \\ 
& $-16.6$ & $-1.53$ & $-2.60$ \\ \hline
{\small $R_i$ [yr$^{-1}$]}  & $130_{-29}^{+59}$ & $30_{-4.6}^{+4.9}$ & $60_{-4.6}^{+8.0}$ \\ \hline
\end{tabular}
\caption{\label{tab:upperlimit}Summary of the upper limit to the Galactic event rate. The errors for the
upper limit are evaluated in section VI in detail. }
\end{table}

\section{Statistical and systematic errors}


We consider various error sources which affect the detection probability. 
These are summarized in Table \ref{tab:errors}. 

\begin{table}[htbp]
\begin{tabular}{cccc}
\hline
& DT6 & DT8 & DT9 \\ \hline\hline
Threshold    & $+0.001$ & $+0.031$ & $+0.013$ \\
                    & $-0.000$  & $-0.024$  & $-0.022$  \\  \hline
Monte Carlo & $\pm 0.093$ & $\pm 0.014$ & $\pm 0.080$ \\ \hline
Calibration  & $+0.034$ & $+0.045$ & $+0.040$ \\ 
                    & $-0.028$  & $-0.041$ & $-0.039$ \\ \hline
$(\delta\epsilon_i)_{\rm fluct}$ 
& $+0.035$ & $+0.056$ & $+0.042$ \\
& $-0.029$ & $-0.049$ &  $-0.045$ \\ \hline\hline
Wave form  & $-0.028$ & $-0.041$   & $-0.039$ \\ \hline
Binary distribution &  $\pm0.028$ & $\pm0.032$ & $\pm0.031$ \\
model                     &  &  &  \\ \hline 
$(\delta\epsilon_i)_{\rm model}$ 
& $+0.028$ & $+0.032$ & $+0.031$ \\
& $-0.056$ & $-0.073$ &  $-0.070$ \\ \hline\hline
\end{tabular}
\caption{\label{tab:errors} The various error sources and their value in the detection probability.}
\end{table}

\noindent
\underline{\it Threshold} ~ 
The method to derive the upper limit to the event rate in this paper requires the evaluation of 
the threshold which corresponds to a given false alarm rate. 
This is done by fitting the distribution of triggers as explained in section V. 
There are statistical errors of the fitting due to the fluctuation of 
the number of background triggers. 
This error results in the error of  the threshold, and the detection probability. 
We have the error of $-0.02 \sim +0.03$ 
in the detection probability in DT8 and DT9 cases. 
In DT6 case, as explained in Section V, 
the error of the fitting due to the non-linear 
property of the distribution was already incorpolated in the fitting.  
The statistical error of the fitting itself was very small in DT6 case, $\lesssim 10^{-3}$.
\\
\underline{\it Monte Carlo} ~ 
The error due to the Monte Carlo injectioin test with a limited trial number is 
given by  $\sqrt{\epsilon_i(1-\epsilon_i)/n_i}$ where $n_i$ is the number 
of the Monte Carlo trials of each run.  This Monte Carlo error becomes about $\pm 0.01$ in the detection 
probability. 
\\
\underline{\it Calibration} ~ 
The calibration of the sensitivity of TAMA300 is done by monitoring 
continuously the response of an injected sinusoidal test signal. 
The error of this response is much smaller than the normalization error discribed below, 
and can be neglected here.  
In the determination of the transfer function, 
there are two possible effects which affect the calibration uncertainty. 
One is an overall normalization error associated with the magnetic actuation strength 
uncertainty and its effect on calibration, and the other is uncertainty in the frequency-dependent
response. Although the error in the normalization is of order 5\%, the long-term drift is unknown.
We thus conservatively adopt 10\%. 
The frequency-dependent error is known to be much less than 
10\%, and thus it is absorbed in the uncertainty in the normalization.
The calibration uncertainty  leads to the errors of $-0.03\sim +0.05$ in the detection probability. 
This calibration error is expected to depend on the different observation runs, and 
is expected to drift and/or fluctuate even within an observation runs.
\\
\underline{\it Binary distribution model} ~ 
We have adopted a specific model for the distribution of binary neutron stars 
in our Galaxy. If the distance between the Sun and the galactic center 
is different from the adopted value, the detection probability will be changed. 
The uncertainty of this distance $\pm 0.9$kpc leads to the uncertainty of 
the detection probability about $\pm 0.03$. \\
\underline{\it Wave form} ~ 
We used the wave form based on the 2.5 PN order. However, currently the best template has 
3.5PN order. The uncertainty of $\rho$ due to this is at most 6\%. 
However, it is reported that the PN wave form itself may contain uncertainty \cite{ref:PNun}. 
We thus adopt 10\% reduction of the estimated $\rho$ as an uncertainty. 
This produces an error of $-0.03\sim -0.04$ in the detection probability. 

The above errors propagate to the upper limit of the event rate for each run. 
We take a quadratic sum of the errors due to threshold, Monte Carlo injection test, and calibration,
since they show the property of fluctuation, and since they are independent each other. 
The sum of these errors, $(\delta \epsilon_i)_{\rm fluct}$, is listed in Table \ref{tab:errors}. 
The errors due to the binary distribution model and the theoretical wave form
produce, simply, the shift of  the detection probability. We thus take a linear sum of them 
conservatively.  The sum of these errors, $(\delta \epsilon_i)_{\rm model}$ is listed in Table \ref{tab:errors}. 
We denote the effect of these errors to the upper limit for each run, 
$R_i$, as $(\delta R_i)_{\rm fluct}$ and $(\delta R_i)_{\rm model}$ respectively,
which are shown in Table \ref{tab:upperlimit}. 
When we evaluate the total error for each run, 
we take a quadratic sum of $(\delta R_i)_{\rm fluct}$ and $(\delta R_i)_{\rm model}$,
since they are independent each other. 
As shown in Table \ref{tab:upperlimit}, 
the  errors of the upper limit to the event rate for each run
become $+59/-29$ [yr$^{-1}$] for DT6, $+4.9/-4.6$ [yr$^{-1}$] for DT8, 
and $+8.0/-4.6$ [yr$^{-1}$] for DT9. 

Finally, we evaluate the error for the combined upper limit, Eq.(\ref{eq:combinedupperlimit}).
The effect of $(\delta \epsilon_i)$ to $R$ is evaluated by taking a quadratic 
sum of each effect of $(\delta \epsilon_i)$ to $R$, and we have $+0.965/-1.08$ [yr$^{-1}$]. 
The effect of $(\delta \epsilon_i)_{\rm model}$ to $R$ is evaluated by simply shifting 
each $\epsilon_i$ in Eq.(\ref{eq:combinedupperlimit}), and we obtain
$+2.86/-1.05$ [yr$^{-1}$].  
The total error in $R$ is evaluated by taking a quadratic sum of these two errors. 
We have  the upper limit with error, 
$R=17^{+3.05}_{-1.51} [{\rm yr}^{-1}]$. 
By taking larger value as a conservative upper limit, we obtain
\begin{eqnarray}
R=20 ~[{\rm yr}^{-1}].
\end{eqnarray}
This value is much larger than an astrophysically expected value, $8.3\times 10^{-5}$[yr$^{-1}$]
\cite{ref:Kalogera} for the coalescence of neutron star binaries. 
However, this rate is smaller than that obtained by LIGO S2 search, 47 ~[yr$^{-1}$MWEG$^{-1}$],
or by LIGO-TAMA joint analysis, 49~[yr$^{-1}$MWEG$^{-1}$]. 
Main reason for this is that the length of data used in our analysis is much longer than these
analyses. 

\section{Summary and Discussion}

In this paper, we have presented the results from the TAMA300 data analysis 
to search for gravitational waves from inspiraling compact binaries
in a mass range, $1-3 M_\odot$. 
We analyzed DT4, DT5, DT6, DT8, and DT9 data of TAMA300. 
There were no triggers which deviate from the tail of the distribution 
of triggers significantly. We thus conclude that 
there is no candidate trigger which can be interpreted as a real gravitational signal. 
By using the long and sensitive data from DT6, DT8 and DT9, 
we obtained upper limits to the Galactic event rate from each observation run. 
We combined these results and obtained a single upper limit, 
20 [yr$^{-1}$] at a 90\% confidence level from these three observation runs. 
We evaluated the systematic errors due to various effects 
such like the uncertainty of calibration and the uncertainty of the background estimation. 
In the upper limit, these effects are included. 

The upper limit obtained in this paper is much larger than an astrophysically 
expected value for the coalescence of neutron star binaries.
However, this upper limit is significant since it is derived by observation. 
Nevertheless, more sensitive detectors are necessary to obtain more 
stringent upper limit to the event rate, and to detect the signal. 
TAMA300 is now improving the suspension system by installing the
Seismic Attenuation System in order to obtain better sensitivity and better stability. 
When it is finished, it is expected to have much better sensitivity than DT9.
LIGO has already been performed 3rd and 4th scientific runs 
with better sensitivity than S2. 
Further, LIGO is now conducting the 5th scientific run since November 2005, 
with its design sensitivity. It can detect the inspiraling binaries up to $\sim$10Mpc distance. 
They are expected to be able to set a much more stringent upper limit. 

When the spin angular momentum of compact objects cannot be neglected,
the spinless template is not good enough to detect the signal, 
and we need to employ templates with spins. 
However, since the number of parameters becomes much larger than 
the non-spinning case, it requires very powerful computer resources. 
One way to avoid the use of the full templates with spins will be to use 
some phenomenological templates with small number of parameters \cite{ref:BCV2}.
We will work on such cases in the future. 

Despite of the improvement and long term observation of current detectors, 
the chance to detect gravitational waves by these first generation detectors 
will not be very large. 
We need more sensitive detectors, such like 
advanced LIGO \cite{ref:advLIGO} and LCGT \cite{ref:LCGT}. 
These detectors will detect gravitational waves
frequently, and will be used to investigate the strong field region of gravity 
and the astrophysics of compact objects. 

\section{Acknowledgments}

This work was supported in part by a Grant-in-Aid for Scientific Research on Priority Area (415)
of the Ministry of Education, Culture, Sports, Science and Technology of Japan.

\end{document}